\documentclass[twoside]{article}
\usepackage[section]{algorithm}
\usepackage{amsthm,amsmath,amssymb,amsmath,amsfonts,graphicx,fancyhdr,algorithmic,longtable,listings,color}

\def\enddemo{\qed \endtrivlist}
\expandafter\let\csname enddemo*\endcsname=\enddemo

\def\qedsymbol{\ifmmode\bgroup\else$\bgroup\aftergroup$\fi
  \vcenter{\hrule\hbox{\vrule
height.6em\kern.6em\vrule}\hrule}\egroup}
\def\qed{\ifmmode\else\unskip\nobreak\fi\quad\qedsymbol}

 \newtheorem{thm}{Theorem}[section]

 \newtheorem{exm}{Example}[section]

\newcommand{\ra}[1]{\mathrm{rank}({#1})}  

 \font\ssr=cmss8 \font\sst=cmtt8 
 \font\ssi=cmti8

\title{\bf About the generalized $LM$-inverse and\\
the Weighted Moore-Penrose inverse}
\begin{footnotesize}
\author{\frenchspacing
\bf Milan B. Tasi\' c\footnote{Corresponding author\ }\ , Predrag S. Stanimirovi\' c,\ Selver H. Pepi\' c\\
{\ssi University of Ni\v{s}, Faculty of Sciences and Mathematics,}\\
{\ssi Vi\v segradska 33, 18000 Ni\v s, Serbia}
\footnote{The authors gratefully acknowledge support from the research project 144011 of the Serbian Ministry of Science.}\\
{\ssi E-mail:} {\sst milan12t@ptt.rs},\ \ {\sst pecko@pmf.ni.ac.rs},\ \ {\sst p\_selver@yahoo.com}\\
}
\end{footnotesize}
\date{}
\begin{document}

\maketitle

\begin{abstract}
The recursive method for computing the generalized $LM$- inverse of a constant rectangular matrix augmented by
a column vector is proposed in \cite{Udwadia,Udwadia2}.
The corresponding algorithm for the sequential determination of the generalized $LM$-inverse is established in the present paper.
We prove that the introduced algorithm for computing the generalized $LM$ inverse and
the algorithm for the computation of the weighted Moore-Penrose inverse developed by Wang in \cite{Wang}
are equivalent algorithms.
Both of the algorithms are implemented in the present paper using the package {\ssr MATHEMATICA}.
Several rational test matrices and randomly generated constant matrices are tested and the CPU time is compared and discussed.

\frenchspacing \itemsep=-1pt
\begin{description}
\item[] AMS Subj. Class.: 15A09, 68W30.
\item[] Key words: Generalized inverses, LM-inverse, Weighted Moore-Penrose inverse, rational matrices, {\ssr MATHEMATICA\/},
Partitioning method.
\end{description}
\end{abstract}

\definecolor{listinggray}{gray}{0.9}

\section{Introduction}

As usual, let $\Bbb C$ be the set of complex numbers, $\Bbb C^{m\times n}$
be the set of $m \times n $ complex matrices, and
${\Bbb C}^{m\times n}_r\!=\!\{X\in {\Bbb C}^{m \times n}\, :\,\,\, \ra{X}\!=\!r\}$.
For any matrix $A\in\Bbb {C}^{m \times n}$ and positive definite
matrices $M$ and $N$ of the orders $m$ and $n$ respectively, consider the following
equations in $X$, where $*$ denotes conjugate and transpose:
$$\begin{array}{ll}
  (1)\qquad \   AXA=A & (2)\qquad XAX\! =\! X \\
  (3M)\quad  (MAX)^*=MAX & (4N)\quad (NXA)^*\! =\! NXA.
\end{array}$$
The matrix $X$ satisfying equations $(1)$, $(2)$, $(3M)$ and $(4N)$ is called the weighted Moore-Penrose inverse of  $A$,
and it is denoted by $X=A_{M,N}^{\dagger}$.
Especially, in the case $M=I_m$ and $N=I_n$, the matrix $X=A_{M,N}^{\dagger}$ becomes
the Moore-Penrose inverse of $A$, and it is denoted by $X=A^{\dagger}$.

\smallskip
Various methods for computing the Moore-Penrose inverse of a matrix are known.
The main methods are based on the Cayley-Hamilton theorem, the full-rank factorization
and the singular value decomposition (see the example \cite{BIG}).
The Greville's {\it partitioning method\/}, introduced in \cite{Gre}, is one of the most
efficient algorithms for computing the Moore-Penrose inverse.
Two different proofs for the Greville's method were presented in \cite{Campbell,Wang1}.
Udwadia and Kalaba gave an alternative and a simple constructive proof of Greville's formula in \cite{Udwadia2N}.
In \cite{Fan} Fan and Kalaba determined the Moore-Penrose inverse of matrices
using dynamic programming and the Belman's principle of optimality.
Sivakumar in \cite{Sivakumar} used the Greville's formula for $A_k^\dagger =[A_{k-1}\, | a_k]^\dagger $
and just verified that it satisfies the four Penrose equations.
This provides a proof of the Greville's method by the verification.

\smallskip
The Greville's algorithm is used in various computations, where its dominance is verified over various direct met\-hods
for the pseudoinverse computation.
The computational experience presented in \cite{Layton} is:
"When applied to a square, fully populated, non-symmetric case, with independent
columns, the Greville's algorithm was found that the approach can be up to 8 times faster than the conventional approach of using
the SVD; rectangular cases are shown to yield similar levels of speed increase".
The Greville's method has been used as a benchmark for the calculation of the pseudo-inverse.

\smallskip
Due to its computational dominance, this method has been extensively
applied in many mathematical areas, such as statistical inference, filtering theory, linear estimation theory, optimization
and more recently analytical dynamics \cite{Udwadia3N} (see also \cite{Kurmayya}).
An application in a direct approach for computing the gradient of the pseudo-inverse is presented in \cite{Layton}.
It has also found wide applications in database and the neural network computation \cite{Mohi}.
In the paper \cite{Itiki}, the sequential determination of the Moore-Penrose inverse by dynamic programming is applied
to the diagnostic classification of electromyography signals.

\smallskip
There is a lot of extensions of the partitioning method.
Wang in \cite{Wang} generalized Greville's method to the weighted Moore-Penrose inverse.
Also, the results in \cite{Wang} are proved by using a new technique.
Udwadia and Kalaba developed the recursive relations for the different types of generalized inverses \cite{UdwadiaN,UdwadiaJOTA}.
Finally, the Greville's recursive principle is generalized to various subsets of outer inverses and extended to
the set of the one-variable rational and polynomial matrices in \cite{Tasic2}.

\smallskip
The algorithm for the computation of the Moore-Penrose inverse of the one-variable polynomial and/or rational matrix,
based on the Greville's partitioning algorithm, was introduced in \cite{Stanimirovic2}.
The extension of results from \cite{Stanimirovic2} to the set of the two-variable rational and polynomial
matrices is introduced in the paper \cite{Stanimirovic3}.

\smallskip
The Wang's partitioning method from \cite{Wang}, aimed in the computation of the weighted Moore-Penrose inverse, is extended to the set of
the one-variable rational and polynomial matrices in the paper \cite{Tasic1}. Also the efficient algorithm
for computing the weighted Moore-Penrose inverse, appropriate for the polynomial matrices where only a few polynomial coefficients are nonzero,
is established in \cite{Petkovic}.

\smallskip
In the paper \cite{Kurmayya} the authors derived
a formula for the computation of the Moore-Penrose inverse of $M^*M$ and obtained sufficient conditions for its nonnegativity,
where $M = [A\, |\, a]$.

\smallskip
On the other side, there are a few articles which are interested in with computation of the generalized $LM$-inverse.
The definition of the $LM$-inverse and the recursive algorithm of the Greville's type
(for a matrix augmented by a column vector) are given in \cite{Udwadia,Udwadia2}.
The recursive relations in \cite{Udwadia,Udwadia2} are proved by direct verification of the four conditions of the generalized
$LM$-inverse.
Also, these formulae are particularized to obtain recursive relations for the generalized $L$-inverse
of a general matrix augmented by a column \cite{Udwadia2}.
The recursive relations for the determination of the generalized Moore-Penrose $M$-inverse are derived in \cite{Udwadia3}.
Separate relations for the situations when the rectangular matrix is augmented by a row vector and when such a matrix
is augmented by a column vector are considered in \cite{Udwadia3}.
The alternative proof for the determination
of the generalized Moore-Penrose $M$-inverse of a matrix through the direct
verification of the four properties of the Moore-Penrose $M$-inverse are presented in \cite{Phoho}.

\smallskip
It is not difficult to verify that the conditions which characterize the generalized $LM$-inverse are
equivalent with the corresponding equations characterizing the weighted Moore-Penrose inverse.
Moreover, the matrix norms minimization used in (3) and (4) in the article \cite{Udwadia}
also characterizes the weighted Moore-Penrose inverse.
Therefore, the generalized $LM$-inverse and the weighted Moore-Penrose inverse are identical.
In the present paper we compare the corresponding algorithms.
It is realistic to predict that algorithm for computing the weighted Moore-Penrose inverse from \cite{Wang}
and the algorithm for the computation of the generalized $LM$-inverse, introduced in the present paper and based
on the results from \cite{Udwadia,Udwadia2} are the same.
Verification of this prediction is the main result of the present paper.
Therefore, the present paper is continuation of the papers \cite{Petkovic,Stanimirovic2,Tasic1,Udwadia,Udwadia2}.

\smallskip
The structure of the present paper is as follows.
In the second section we restate
the representation of the generalized $LM$-inverse from \cite{Udwadia,Udwadia2}
as well as the representation and algorithm for computing the weighted Moore-Penrose from \cite{Wang}.
We also introduce an effective algorithm for construction of the generalized $LM$-inverse directly using
its representation proposed in \cite{Udwadia,Udwadia2}.
In the third section we provide a proof that two algorithms from the second section are equivalent.
Implementation of both algorithms and a few illustrative examples are presented.

\section{Preliminaries and motivation}

The recursive determination of the weighted Moore-Penrose inverse $A_{M,N}^{\dagger}$
is established in \cite{Wang}.

\smallskip
Let $A\in\Bbb {C}^{m \times n}$ and ${A}_k$ be the submatrix of $A$
consisting of its first $k$ columns. For $k=2,\ldots ,n$ the matrix $A_k$ is partitioned as
\begin{equation}\label{dva1}
A_k=\left[ A_{k-1}\ |\ a_k \right],
\end{equation}
where $a_k$ is the $k$-th column of $A$.

\begin{thm}[G.R. Wang, Y.L. Chen \cite{Wang}] \label{thw}
Let $A\in\Bbb {C}^{m \times n}$ and ${A}_k$ be the submatrix of $A$
consisting of its first $k$ columns. For $k=2,\ldots ,n$ the matrix $A_k$ is partitioned as in $(\ref{dva1})$,
and the matrix $N_k\in\Bbb {C}^{k \times k}$ is the leading principal submatrix of $N$, and $N_k$ is partitioned as

\begin{equation}\label{dva2}
    N_k=\left[
               \begin{array}{cc}
                 N_{k-1} & l_k \\
                 l_k^* & n_{kk} \\
               \end{array}
             \right].
\end{equation}
Let the matrices $X_{k-1}$  and $X_{k}$ be defined by

\begin{equation}
X_{k-1}=(A_{k-1})^{\dagger}_{M,N_{k-1}}, \quad X_k=(A_k)^{\dagger}_{M,N_k},
\end{equation}

\noindent the vectors $d_k$, $c_k$ be defined by
\begin{eqnarray}
  d_k &=& X_{k-1}a_k\label{dva5} \\
  c_k &=& a_k-A_{k-1}d_k = \left( I-A_{k-1}X_{k-1}\right) a_k.\label{dva6}
\end{eqnarray}

Then
\begin{equation}\label{dva4}
    X_{k}\!\!=\!\!\left[
          \begin{array}{c}
            X_{k-1}\!-\!\left( d_k+(I-X_{k-1}A_{k-1}\right)   N_{k-1}^{-1}l_k)b_k^* \\
            b_k^* \\
          \end{array}
        \right],
\end{equation}
where
\begin{equation}\label{dva7}
    b_k^*=\left\{
       \begin{array}{ll}
             \left( c^*_k M c_k\right) ^{-1}c_k^* M, & c_k\ne 0 \\\\
             \delta_k^{-1}\left( d_k^* N_{k-1}-l_k^*\right) X_{k-1}, & c_k=0,
       \end{array}
     \right.
\end{equation}
and
$\delta_k=n_{kk}+d^*_k N_{k-1} d_k-\left( d^*_k l_k+l_k^* d_k(s)\right)-l_k^*\left( I-X_{k-1} A_{k-1}\right) N_{k-1}^{-1}l_k.$
\end{thm}

According to the above theorem the next Algorithm \ref{algwang} is introduced in \cite{Wang}.

\vfill\eject
\begin{algorithm}
\caption{Computing the weighted M-P inverse $A_{M,N}^\dag$ using algorithm from \cite{Wang}.}
\begin{algorithmic}[1]
\REQUIRE Let $A\in \Bbb{C}^{m\times n}$, $M$ and $N$ be p.d. matrices of the order $m$ and $n$ respectively.
\STATE $A_1=a_1.$
\IF {$a_1=0,$}
\STATE $X_1=(a_1^* M a_1)^{-1} a_1^* M;$
\ELSE
\STATE $X_1=0.$
\ENDIF
\FOR{$k=2$ to $n$}
\STATE $d_{k} \!=\!X_{k-1}a_k,$
\STATE $c_k=a_k-A_{k-1}d_k,$
\IF {$c_k\ne0,$}
\STATE $b_k^*=(c_k^* M c_k)^{-1} c_k^* M, \ {\bf goto} \ Step\ \ref{sest},$
\ELSE
\STATE $\delta_k=n_{kk}+d_k^* N_{k-1} d_k - (d_k^* l_k+l_k^* d_k)-l_k^*(I-X_{k-1}A_{k-1})N_{k-1}^{-1} l_k,$
\STATE $b_k^*=\delta_k^{-1}(d_k^*N_{k-1} -l_k^*) X_{k-1},$
\ENDIF
\STATE $X_k=\left[\begin{array}{cc} X_{k-1}-(d_k+(I-X_{k-1}A_{k-1})N_{k-1}^{-1} l_k)b_k^*\\
b_k^*\end{array}\right].\label{sest}$
\ENDFOR
\STATE {\bf return} $A_{MN}^\dag=X_n.$
\end{algorithmic}\label{algwang}
\end{algorithm}

Also, the next auxiliary Algorithm \ref{algwang1}, required in Algorithm \ref{algwang} is stated in \cite{Wang}.

\begin{algorithm}
\caption{Computing the inverse matrix $N^{-1}$.}
\begin{algorithmic}[1]
\REQUIRE Let $N_{k}=\left[\begin{array}{cc} N_{k-1}&l_k\\l_k^*& n_{kk}\end{array}\right]\in \Bbb{C}^{k\times k}$ be the leading principal submatrix of p.d. matrix $N$.
\STATE $N_1^{-1}=n_{11}^{-1}.$
\FOR{$k=2$ to $n$}
\STATE $g_{kk} \!=\!(n_{kk}-l_k^* N_{k-1}^{-1} l_{k})^{-1},$
\STATE $f_k=-g_{kk} N_{k-1}^{-1} l_k,$
\STATE $E_{k-1}=N_{k-1}^{-1}+g_{kk}^{-1} f_k f_k^*,$
\STATE $N_{k}^{-1}=\left[\begin{array}{cc} E_{k-1}&f_k\\f_k^*& g_{kk}\end{array}\right].$
\ENDFOR
\STATE {\bf return} $N^{-1}=N_n^{-1}.$
\end{algorithmic}\label{algwang1}
\end{algorithm}

The definition of the generalized $LM$-inverse is given in \cite{Udwadia}, and it is
based on the usage of the linear equation $Ax = b$, where $A$ is an $m\times n$ matrix,
$b$ is an $m$-vector and $x$ is an $n$-vector.
The matrix $A_{LM}^{\dagger}$ is such that the vector $x$, uniquely given by
$x = A_{LM}^{\dagger}b,$ minimizes both of the following two vector norms (conditions (3) and (4) from \cite{Udwadia})
$$
\aligned G &= \parallel L^{1/2}(Ax - b)\parallel^2 = \parallel Ax - b \parallel^2_L,\\
H &= \parallel M^{1/2}x\parallel^2 = \parallel x \parallel^2_M,
\endaligned
$$
where $L$ is an $m\times m$ symmetric positive-definite matrix and $M$ is an $n\times n$ symmetric positive-definite matrix.

\smallskip
The recursive formulae for determining the generalized $LM$-inverse $A_{L,M}^{\dagger}$ of any given matrix $A$
are introduced in \cite{Udwadia,Udwadia2}, and they are restated here for the sake of completeness.

\begin{thm} [F.E. Udwadia, P. Phohomsiri \cite{Udwadia,Udwadia2}]\label{thudw}
The generalized $LM$-inverse of any given matrix $B=[A\ |\ a]\in \Bbb C^{m\times n}$ is determined using the
following recursive relations:
\begin{eqnarray}\label{ujedan}
    B_{L,M}^{\dagger}=[A\ |\ a]_{L,M}^{\dagger}=
\left\{ \begin{array}{ll}
    \left[
          \begin{array}{cl}
            A_{L,M\_}^{\dagger}\!-\!A_{L,M\_}^{\dagger}\, a \, d_L^{\dagger}- p\, \, d_L^{\dagger}\\
            d_L^{\dagger}
          \end{array}
        \right] , & d = \left( I-AA_{L,M\_}^{\dagger}\right) a\neq 0; \\ {} & {} \\
     \left[
          \begin{array}{cl}
            A_{L,M\_}^{\dagger}\!-\!A_{L,M\_}^{\dagger}\, a h- p\, h\\
            h
          \end{array}
        \right], & d = \left( I-AA_{L,M\_}^{\dagger}\right) a=0,
\end{array} \right.
\end{eqnarray}
where $A$ is an $m\times (n-1)$ matrix, $a$ is a column vector of $m$ components,
\begin{equation*}
\aligned
d_L^{\dagger}&=\frac{d^T L}{d^T L d},\ h=\frac{q^TM U}{q^T M q} ,\\
  U  &= \left[\begin{array}{cl}
            A_{L,M\_}^{\dagger}\\
            0_{1\times m}
          \end{array}\right],\
  q =\left[\begin{array}{cl}
            v+p\\
            -1
          \end{array}
        \right],\
v=A_{L,M\_}^{\dagger}a
\endaligned
\end{equation*}
\noindent and
$$p=\left( I-A_{L,M\_}^{\dagger}A\right) M^{-1}_{\_}\widetilde{m}.$$
Note that $L$ is a symmetric positive definite $m\times m$ matrix, and
\begin{equation}\label{PartT}
    M=\left[
               \begin{array}{ll}
                 M_{\_} & \widetilde{m} \\\\
                 \widetilde{m}^T & \bar{m}
               \end{array}
             \right],
\end{equation}
where $M$ is a symmetric positive-definite $n\!\times\! n$ matrix,
$M_{\_}$ is a symmetric positive-definite $(n -\! 1)\times (n - 1)$ matrix, $\widetilde{m}$ is a column vector of $n-1$ components,
and $\bar{m}$ is a scalar.
\end{thm}

Theorem \ref{thudw} assumes in $B=[A\ |\ a]$ that the matrix $B$ is obtained augmenting the matrix $A$ by an appropriate column vector $a$.
In the rest of the paper we assume that $B=[A\ |\ a]$ is just the partitioning $(\ref{dva1})$: $B=A_k,\ A=A_{k-1},\ a=a_k$.
Moreover, it is clear that the following notations immediately follows from Algorithm \ref{algwang}:
$$B_{L,M}^{\dagger}=X_k,\ A^\dagger =X_{k-1}.$$
Also, we use the following denotation for the matrix $M$ defined in \eqref{PartT}:
\begin{equation}
    M=\left[
               \begin{array}{ll}
                 M_{k-1} & \widetilde{m}_k \\\\
                 \widetilde{m}_k^T & m_{k,k}
               \end{array}
             \right],
\end{equation}
Finally, the vector $d$ corresponding to the first $k$ columns of $A$ is denoted by $d_k$.

\smallskip
According to the above Theorem \ref{thudw} we introduce the next algorithm.

\vfill\eject
\begin{algorithm}
\caption{Computing the $LM$-inverse $A_{LM}^\dag$ using the representation from \cite{Udwadia}.}
\begin{algorithmic}[1]
\REQUIRE Let $A\in \Bbb{C}^{m\times n}$, $L$ and $M$ be p.d. matrices of order $m$ and $n$ respectively.
\STATE $A_1=a_1.$
\IF {$a_1=0$}
\STATE $X_1=(a_1^* L a_1)^{-1} a_1^* L$
\ELSE
\STATE {$X_1=0.$}
\ENDIF
\FOR{$k=2$ to $n$}

\STATE $d_k=\left( I-A_{k-1}X_{k-1}\right)a_k$,
\STATE $p=\left( I-X_{k-1}A_{k-1}\right) M_{k-1}^{-1}\widetilde{m}_k$,
\IF {$d_k\ne0$}
\STATE $b_k^*=\frac{d_k^* L}{d_k^* L d_k}\  {\bf goto} \ Step \ \ref{deset}$
\ELSE
\STATE $q=\left[\begin{array}{cl}
            X_{k-1}a_k+p\\
            -1 \\
          \end{array}
        \right]$,
\STATE  $U =\left[\begin{array}{cl}
            X_{k-1}\\
            0 \\
          \end{array}\right]$,
\STATE $\ b_k^*=\frac{q^*}{q^* M_{k} q} M_{k} U$,
\ENDIF
\STATE $X_k=\left[\begin{array}{cc} X_{k-1}-X_{k-1}a_k b_k^* -p\ b_k^*\\
b_k^*\end{array}\right].$\label{deset}
\ENDFOR
\STATE {\bf return} $A_{L,M}^\dag=X_n.$
\end{algorithmic}\label{algudwa}
\end{algorithm}

\medskip
It is clear from restated definitions that the generalized $LM$-inverse is just the weighted Moore-Penrose inverse.
Therefore, Algorithm \ref{algudwa} and algorithms \ref{algwang}, \ref{algwang1} together produce identical result - the
weighted Moore-Penrose inverse of the given $m\times n$ matrix.
In the next section we compare the described algorithms.

\section{Comparison of algorithms}

\begin{thm} 
Algorithm \ref{algudwa} is equivalent to algorithms \ref{algwang} and \ref{algwang1}.
\end{thm}

\begin{demo}
In order to ensure unambiguous, during the proof we assume that the symbol $W$ in a superscript denotes terms from the Wang's algorithm;
similarly we use the convention that $U$, as a superscript, denotes terms from the Udwadia's algorithm, elsewhere it is necessary.
Since the $LM$-inverse is just the weighted Moore-Penrose inverse, we conclude that the matrix $M$ in Algorithm \ref{algwang}
is just the matrix $L$ in Algorithm \ref{algudwa} and the
matrix $N$ in Algorithm \ref{algwang} is analogous with the matrix $M$ in Algorithm \ref{algudwa}.
Therefore, it is not necessarily to mark the matrices $L,M,N$ and $A$ by appropriate superscript.

\smallskip
We prove the theorem by verifying the equivalence of the outputs from the corresponding algorithmic steps of mentioned algorithms.
The proof proceeds by the mathematical induction.

\smallskip
The proof for the case $k=1$ in view of Step 3 in both algorithms is trivial.
Assume that the statement is valid for the first $k-1$ columns, i.e.
\begin{equation}\label{Pozovi}
X_{k-1}^U=X_{k-1}^W=(A_{k-1})^\dagger_{M,N_{k-1}}=A_{LM_{\_}}^{\dagger}.
\end{equation}

Now we verify the inductive step.
Wang used the matrix $X_k$ in the form
\begin{equation}
    X_{k}^{W}\!\!=\!\!\left[
          \begin{array}{c}
            X_{k-1}^{W}\!-\!\left( d_k^{W}+(I-X_{k-1}^{W}A_{k-1}\right)   N_{k-1}^{-1}l_k)(b_k^*)^{W} \\
            (b_k^*)^{W} \\
          \end{array}
        \right],
\end{equation}
\noindent while Udwadia observed two cases, as in $(\ref{ujedan})$.

\smallskip
If we denote with
\begin{equation}
    (b_{k}^{*})^{U}\!\!=\!\!\left\{
          \begin{array}{c}
            d_L^{\dag},\quad d_k^{U}\ne 0\\ \smallskip
            h,\quad  d_k^{U}=0,
          \end{array}
        \right.
\end{equation}
then the equalities in (\ref{ujedan}) become
\begin{equation}
    X_{k}^{U}\!\!=\!\!\left[
          \begin{array}{c}
            X_{k-1}^{U}\!-\!X_{k-1}^{U}a_k (b_k^*)^{U} - p \, (b_k^*)^{U}\\
            (b_k^*)^{U} \\
          \end{array}
        \right].
\end{equation}
Let us show that the output of Step 9 from Algorithm \ref{algwang} is the same as the output of Step 8 from Algorithm \ref{algudwa}:
\begin{eqnarray*}
c_k^{W}\equiv a_k-A_{k-1}d_k^{W}=a_k-A_{k-1}X_{k-1}^{W}a_k=\left( I-A_{k-1}X_{k-1}^{W}\right)a_k\equiv d_k^{U}.
\end{eqnarray*}
Now we show that Step 11 from Algorithm \ref{algwang} and Step 11 from Algorithm \ref{algudwa} are equivalent.
As it is stated above $c_k^{W}=d_k^{U}$, so that in the case $c_k^{W}\neq 0$ we have
$$(b_k^*)^{W}=((c_k^*)^{W} M c_k^{W})^{-1} (c_k^*)^{W} M=((d_k^*)^{U} L d_k^{U})^{-1} (d_k^*)^{U} L=(b_k^*)^{U}.$$
In a similar way it can be verified that Step 14 from Algorithm \ref{algwang} is equivalent to Step 15 from Algorithm \ref{algudwa}.
In the case $c_k^{W}=0$ we can start from the statement in Step 15 from Algorithm \ref{algudwa}.
$$(b_k^*)^{U}=(q^* M_k U)/(q^* M_k q).$$

\noindent From Step 9 of Algorithm \ref{algudwa} and the inductive hypothesis the following holds
$$p=\left(I-X_{k-1}^{U}A_{k-1}\right)N_{k-1}^{-1}l_k=\left(I-X_{k-1}^{W}A_{k-1}\right)N_{k-1}^{-1}l_k,$$
so that we derive the following:
\begin{eqnarray*}
q^T M_k U\!\!&=&\!\!\left[(X_{k-1}^{U} a_k +p)^* \quad  |\quad  -1\right]\left[\begin{array}{cc} N_{k-1}&l_k\\ l_k^*& n_{kk}\end{array}\right]
\left[\begin{array}{c} X_{k-1}^{U}\\0 \end{array}\right]\\
&=&\left[(X_{k-1}^{U} a_k +p)^*N_{k-1}-l_k^* \quad  |\quad   (X_{k-1}^{U}a_k+p)^*l_k-n_{kk}\right]
\left[\begin{array}{c} X_{k-1}^{U}\\0 \end{array}\right]\\
&=&(X_{k-1}^{U} a_k +p)^* N_{k-1}X_{k-1}^{U}-l_k^* X_{k-1}^{U}\ \{\textrm{since } \ X_{k-1}^{U} a_k= X_{k-1}^{W} a_k= d_k^{W}\} \\
&=&(d_k^*)^{W} N_{k-1}X_{k-1}^{W}-l_k^* X_{k-1}^{W}+p^*N_{k-1}X_{k-1}^{W}\\
&=&\left( (d_k^*)^{W} N_{k-1} -l_k^*\right)  X_{k-1}^{W}+\left( \left(I-X_{k-1}^{W}A_{k-1}\right)N_{k-1}^{-1}l_k\right)^* N_{k-1}X_{k-1}^{W}\\
&=&\left( (d_k^*)^{W} N_{k-1} -l_k^*\right) X_{k-1}^{W}+\left( l_k^* N_{k-1}^{-1}\left(I-X_{k-1}^{W}A_{k-1}\right)^*\right) N_{k-1}X_{k-1}^{W}\\
&=&\left( (d_k^*)^{W} N_{k-1} -l_k^*\right) X_{k-1}^{W}+l_k^*X_{k-1}^{W}-l_k^* N_{k-1}^{-1}\left(X_{k-1}^{W}A_{k-1}\right)^* N_{k-1}X_{k-1}^{W}.
\end{eqnarray*}
Since $N_{k-1}$ is the symmetric positive definite applying equality $(4N)$ together with $(\ref{Pozovi})$ the following holds
$$\left(X_{k-1}^{W}A_{k-1}\right)^*N_{k-1}=\left(N_{k-1}X_{k-1}^{W}A_{k-1}\right)^*=N_{k-1}X_{k-1}^{W}A_{k-1},$$
and later
\begin{eqnarray*}
q^T M_k U\!\!&=&\!\! \left( (d_k^*)^{W} N_{k-1} -l_k^*\right) X_{k-1}^{W}+l_k^*X_{k-1}^{W}-l_k^* N_{k-1}^{-1}N_{k-1}X_{k-1}^{W}A_{k-1}X_{k-1}^{W}\\
&=&\left((d_k^*)^{W} N_{k-1} -l_k^*\right) X_{k-1}^{W}+l_k^* \left(X_{k-1}^{W}-X_{k-1}^{W}A_{k-1}X_{k-1}^{W}\right)\\
&=&\left((d_k^*)^{W} N_{k-1} -l_k^*\right) X_{k-1}^{W}+l_k^* \left(X_{k-1}^{W}-X_{k-1}^{W}\right)\\
&=&\left( (d_k^*)^{W} N_{k-1} -l_k^*\right) X_{k-1}^{W}.
\end{eqnarray*}

Moreover, we have
\begin{eqnarray*}
q^T M_k q\!\!&=&\!\!\left[(X_{k-1}^{U} a_k +p)^* \quad  |\quad  -1\right]\left[\begin{array}{cc} N_{k-1}&l_k\\ l_k^*& n_{kk}\end{array}\right]
\left[\begin{array}{c} X_{k-1}^{U}a_k+p\\-1 \end{array}\right]\\
\!\!&=&\!\!\left[(X_{k-1}^{U} a_k+p)^* N_{k-1}-l_k^* \quad  |\quad  (X_{k-1}^{U}a_k+p)^* l_k-n_{kk}\right]
\left[\begin{array}{c} X_{k-1}^{U}a_k+p\\-1 \end{array}\right]\\
\!\!&=&\!\!((X_{k-1}^{U} a_k+p)^* N_{k-1}-l_k^*)(X_{k-1}^{U}a_k+p) - (X_{k-1}^{U} a_k+p)^* l_k+n_{kk}\\
\!\!&=&\!\!(X_{k-1}^{U} a_k+p)^* N_{k-1}(X_{k-1}^{U}a_k+p)-l_k^*(X_{k-1}^{U} a_k+p) - (X_{k-1}^{U} a_k+p)^* l_k+n_{kk}\\
\!\!&\phantom{=}&\!\!\ (\textrm{since } \ \{  X_{k-1}^{U} a_k= X_{k-1}^{W} a_k= d_k^{W}\})\\
\!\!&=&\!\!(d_k^{W}+p)^* N_{k-1}(d_k^{W}+p)-l_k^*(d_k^{W}+p) - (d_k^{W}+p)^* l_k+n_{kk}\\
\!\!&=&\!\! (d_k^*)^{W} N_{k-1}d_k^{W}-l_k^* d_k^{W}-(d_k^*)^{W} l_k+n_{kk}-l_k^*p\\
\!\!&\phantom{=}&\ \ +\, p^* N_{k-1}d_k^{W}+(d_k^*)^{W}N_{k-1}p+ p^* N_{k-1}p-p^* l_{k}.
\end{eqnarray*}
Furthermore
\begin{equation}\label{suma}
p^* N_{k-1}d_k^{W} + (d_k^*)^{W}N_{k-1}p+ p^* N_{k-1}p-p^* l_{k}=0.
\end{equation}
First we show that $p^* N_{k-1}d_k^{W}=0$, as follows
\begin{eqnarray*}
p^* N_{k-1}d_k^{W}\!\!&=&\!\!
\left( \left(I-X_{k-1}^{W}A_{k-1}\right)N_{k-1}^{-1}l_k\right)^* N_{k-1}X_{k-1}^{W}a_k\\
\!\!&=&\!\!l_k^* N_{k-1}^{-1}\left(I-X_{k-1}^{W}A_{k-1}\right)^* N_{k-1}X_{k-1}^{W}a_k\\
\!\!&=&\!\!l_k^* N_{k-1}^{-1} \left(N_{k-1}-\left(X_{k-1}^{W}A_{k-1}\right)^*N_{k-1}\right) X_{k-1}^{W}a_k\\
\!\!&=&\!\!l_k^* N_{k-1}^{-1} \left(N_{k-1}-N_{k-1}X_{k-1}^{W}A_{k-1}\right) X_{k-1}^{W}a_k\\
\!\!&=&\!\!l_k^* N_{k-1}^{-1}N_{k-1}\left(I-X_{k-1}^{W}A_{k-1}\right) X_{k-1}^{W}a_k\\
\!\!&=&\!\!l_k^* \left(I-X_{k-1}^{W}A_{k-1}\right) X_{k-1}^{W}a_k\\
\!\!&=&\!\!l_k^* \left(X_{k-1}^{W}-X_{k-1}^{W}A_{k-1}X_{k-1}^{W}\right) a_k\\
\!\!&=&\!\!l_k^* \left(X_{k-1}^{W}-X_{k-1}^{W}\right) a_k=0.
\end{eqnarray*}
Also, from the above equality, we have
$$(d_k^*)^{W}N_{k-1}p=(p^* N_{k-1}d_k^{W})^*=0.$$
Finally, the last term of the sum in the left hand side of $(\ref{suma})$, is equal to $p^* N_{k-1}p-p^* l_{k}$,
and it is also equal to zero:
\begin{eqnarray*}
p^* N_{k-1}p-p^* l_{k}\!\!&=&\!\!p^*\left( N_{k-1}\left(I-X_{k-1}^{W}A_{k-1}\right)N_{k-1}^{-1}l_k-l_k\right) \\
\!\!&=&\!\!p^*(-N_{k-1}X_{k-1}^{W}A_{k-1}N_{k-1}^{-1}\, l_k)\\
\!\!&=&\!\!p^*\left(-(X_{k-1}^{W}A_{k-1})^* \, l_k\right)\\
\!\!&=&\!\!\left( \left(I-X_{k-1}^{W}A_{k-1}\right)N_{k-1}^{-1}l_k\right) ^*(-A^*_{k-1} (X_{k-1}^*)^{W}l_k)\\
\!\!&=&\!\!-l_k^*N_{k-1}^{-1}{A^*_{k-1}}(X_{k-1}^*)^{W}l_k+l_k^*N_{k-1}^{-1}A_{k-1}^{*}(X^{*}_{k-1})^{W}A_{k-1}^*(X^{*}_{k-1})^{W}l_k\\
\!\!&=&\!\!-l_k^*N_{k-1}^{-1}{A^*_{k-1}}(X_{k-1}^*)^{W}l_k+l_k^*N_{k-1}^{-1}A_{k-1}^{*}(X^{*}_{k-1})^{W}l_k=0.
\end{eqnarray*}

Continuing the transformation for $q^T M_k q$ we have
\begin{eqnarray*}
q^T M_k q\!\!&=&\!\!(d_k^*)^{W} N_{k-1}d_k^{W}-l_k^* d_k^{W}-(d_k^*)^{W}l_k+n_{kk}-l_k^*\left(I-X_{k-1}^{W}A_{k-1}\right)N_{k-1}^{-1}l_k\\
&=&\delta_k.
\end{eqnarray*}
Now we are able to continue the rest of our proof. According to the equality
$$(b_k^*)^{U}=(q^* M_k U)/(q^* M_k q),$$
Step 15 from Algorithm \ref{algudwa} produces the result
$$(b_k^*)^{U}=\delta_k^{-1}(d_k^*N_{k-1} -l_k^*) X_{k-1}^{W},$$
which is identical to the output $(b_k^*)^{W}$, derived in Step 14 from Algorithm \ref{algwang}.

\medskip
According to  Step 16 of Algorithm \ref{algwang} and Step 17 of Algorithm \ref{algudwa},
the generalized $LM$-inverse and the weighted Moore-Penrose inverse of the first $k$ columns of $A$ are identical:
\begin{equation}
\aligned
    X_k^{U}& =\left[
          \begin{array}{c}
            X_{k-1}^{U}\!-\!\left(X_{k-1}^{U}a_k +p\right)(b_k^*)^{U}\\
            (b_k^*)^{U} \\
          \end{array}
        \right]\\
        & =
\left[ \begin{array}{c}
            X_{k-1}^{W}\!-\!\left(d_k^{W} +\left(I-X_{k-1}^{W}A_{k-1}\right)N_{k-1}^{-1}l_k \right)(b_k^*)^{W}\\
            (b_k^*)^{W} \\
          \end{array}
        \right] \\
        & =X_k^{W}.
\endaligned
\end{equation}
Finally, for the case $k=n$, it immediately follows that $A^\dagger _{L,M}=A^\dagger _{M,N}$, which means that the outputs from both algorithms are
identical.
\end{demo}

\section{Examples}

In order to compare the algorithms from the second section it is necessary to use the precise implementation of the corresponding algorithms.
Details concerning the implementation of the partitioning algorithm corresponding to the
weighted Moore-Penrose inverse can be found in \cite{Tasic1}.
In order to compare the mentioned algorithms we developed a {\ssr MATHEMATICA} code for the implementation of Algorithm \ref{algudwa}.
We later tested results on different types of matrices.
Since the language {\ssr MATHEMATICA} admits symbolic manipulation with data, developed implementations are
immediately applicable to the rational and polynomial matrices.

\begin{exm} \rm Consider the test matrix $A_{11 \times 10}$ from \cite{Zie}, in the case $a=1$
\begin{scriptsize}
$$A=\left[
  \begin{array}{cccccccccc}
    1 &2 &3 &4 &1 &1 &3 &4 &6 &2\\
1 &3 &4 &6 &2 &2 &3 &4 &5 &3\\
2 &3 &4 &5 &3 &3 &4 &5 &6 &4\\
3 &4 &5 &6 &4 &4 &5 &6 &7 &6\\
4 &5 &6 &7 &6 &6 &6 &7 &7 &8\\
6 &6 &7 &7 &8 &1 &2 &3 &4 &1\\
3 &4 &1 &1 &3 &4 &6 &2 &1 &2\\
4 &6 &2 &2 &3 &4 &5 &3 &1 &3\\
4 &5 &3 &3 &4 &5 &6 &4 &2 &3\\
5 &6 &4 &4 &5 &6 &7 &6 &3 &4\\
6 &7 &6 &6 &6 &7 &7 &8 &4 &5 \\
  \end{array}
\right]_{11\times10}$$
\end{scriptsize}
and randomly generated symmetric positive definite matrices $L_{10 \times 10}$ and $M_{11 \times 11}$:
\begin{scriptsize}
$$L=\left[
  \begin{array}{rrrrrrrrrr}
280&-5&-133&-27&-12&-93&-133&-42&84&52\\
-5&216&-93&-23&141&-2&-108&-48&21&165\\
-133&-93&336&1&0&0&9&-5&91&81\\
-27&-23&1&260&-62&-42&-40&6&85&-8\\
-12&141&0&-62&278&-63&-19&-135&9&99\\
-93&-2&0&-42&-63&238&68&-27&-80&-34\\
-133&-108&9&-40&-19&68&290&12&-233&-244\\
-42&-48&-5&6&-135&-27&12&209&-15&-87\\
84&21&91&85&9&-80&-233&-15&332&145\\
52&165&81&-8&99&-34&-244&-87&145&402
  \end{array}
\right]_{10\times 10},$$

$$M\!\!=\!\!\left[
  \begin{array}{rrrrrrrrrrr}
452&91&-186&-97&-161&68&28&16&151&41&-65\\
91&413&-74&-119&-317&-41&-12&67&180&-53&54\\
-186&-74&497&-136&78&-208&-175&-120&9&-99&6\\
-97&-119&-136&371&157&154&129&29&-102&-16&-96\\
-161&-317&78&157&444&28&-39&-201&-165&3&43\\
68&-41&-208&154&28&509&52&55&90&179&-81\\
28&-12&-175&129&-39&52&454&-38&-157&145&22\\
16&67&-120&29&-201&55&-38&408&-9&-16&-100\\
151&180&9&-102&-165&90&-157&-9&257&-32&14\\
41&-53&-99&-16&3&179&145&-16&-32&376&-15\\
-65&54&6&-96&43&-81&22&-100&14&-15&339
  \end{array}
\right]_{11\times 11}.$$
\end{scriptsize}

\smallskip
\noindent The generalized $LM$-inverse $A_{L,M}^\dagger$ from \cite{Udwadia,Udwadia2} and the weighted
Moore-Penrose inverse $A_{M,N}^\dagger$ from \cite{Wang} are both equal to

\smallskip
\begin{scriptsize}
\noindent $\left[
  \begin{array}{rrrrrrrrrrr}
0.755&-0.156&-1.917&0.823&0.143&0.033&0.383&-0.33&0.213&-0.802&0.67\\
-0.542&-0.078&1.683&-0.544&-0.27&0.003&-0.432&0.673&0.075&0.087&-0.347\\
0.346&-0.194&-0.881&0.751&-0.187&0.002&0.32&-0.149&0.711&-1.749&1.003\\
-0.049&0.558&-0.724&0.145&0.065&0.007&0.128&-0.258&-0.245&0.481&-0.141\\
-0.454&-0.013&1.181&-0.79&0.188&0.109&-0.3&0.057&-0.563&1.498&-0.907\\
-1.188&-0.468&2.87&-0.203&-0.792&-0.107&0.233&0.19&1.258&-2.561&1.083\\
0.701&0.364&-2.009&0.492&0.282&0.009&0.395&-0.407&-0.481&0.786&-0.192\\
0.472&0.156&-0.533&-0.675&0.507&-0.02&-0.62&-0.008&-0.822&2.174&-0.848\\
-0.415&-0.481&1.824&-0.22&-0.377&0.002&0.013&0.247&0.615&-1.143&0.257\\
0.328&0.156&-1.388&0.469&0.362&-0.02&0.028&-0.008&-0.47&0.526&-0.2
\end{array}
\right].$
\end{scriptsize}

\smallskip
\noindent The Moore-Penrose inverse can be generated in the case $L=M=\alpha I$, $M=N=\beta I$
\cite{Udwadia}, and it is equal to

\smallskip
\begin{scriptsize}
\noindent $A^\dagger\!=\!\left[
  \begin{array}{rrrrrrrrrrr}
0.294&-0.169&-1.511&1.415&-0.391&0.041&-0.04&-0.26&0.454&-0.264&0.23\\
-0.067&-0.074&1.227&-1.064&0.23&0.001&-0.192&0.649&-0.103&-0.191&-0.102\\
0.227&-0.179&-0.692&0.705&-0.214&-0.009&-0.254&-0.238&1.514&-1.319&0.449\\
-0.165&0.547&-0.657&0.376&-0.116&0.015&0.179&-0.196&-0.456&0.531&-0.104\\
-0.297&-0.01&1.035&-0.972&0.359&0.108&0.238&0.044&-1.065&0.938&-0.366\\
-0.008&-0.474&1.663&-1.319&0.352&-0.103&-0.428&0.224&1.83&-1.844&0.414\\
0.062&0.368&-1.349&1.081&-0.33&0.006&0.426&-0.434&-0.442&0.712&-0.156\\
-0.081&0.158&0.029&-0.144&-0.034&-0.021&0.087&-0.019&-1.499&1.448&-0.138\\
0.195&-0.484&1.198&-0.791&0.211&0.005&-0.19&0.268&0.765&-0.906&0.049\\
-0.081&0.158&-0.971&0.856&-0.034&-0.021&0.087&-0.019&-0.499&0.448&-0.138
\end{array}
\right]$.
\end{scriptsize}
\end{exm}

\begin{exm} \rm Consider the one variable test matrix
\begin{footnotesize}
$$
A=\left[
\begin{array}{llllllllllll}
 x & 1 & 0 & 0 & 0 & 0 & 0 & 0 & 0 & 0 & 0 & 0 \\
 x^2 & x & 1 & 0 & 0 & 0 & 0 & 0 & 0 & 0 & 0 & 0 \\
 x^3 & x^2 & x & 1 & 0 & 0 & 0 & 0 & 0 & 0 & 0 & 0 \\
 x^4 & x^3 & x^2 & x & 1 & 0 & 0 & 0 & 0 & 0 & 0 & 0 \\
 x^5 & x^4 & x^3 & x^2 & x & 1 & 0 & 0 & 0 & 0 & 0 & 0 \\
 x^6 & x^5 & x^4 & x^3 & x^2 & x & 1 & 0 & 0 & 0 & 0 & 0 \\
 x^7 & x^6 & x^5 & x^4 & x^3 & x^2 & x & 1 & 0 & 0 & 0 & 0 \\
 x^8 & x^7 & x^6 & x^5 & x^4 & x^3 & x^2 & x & 1 & 0 & 0 & 0 \\
 x^9 & x^8 & x^7 & x^6 & x^5 & x^4 & x^3 & x^2 & x & 1 & 0 & 0 \\
 x^{10} & x^9 & x^8 & x^7 & x^6 & x^5 & x^4 & x^3 & x^2 & x & 1 & 0 \\
 x^{11} & x^{10} & x^9 & x^8 & x^7 & x^6 & x^5 & x^4 & x^3 & x^2 & x & 1 \\
 x^{12} & x^{11} & x^{10} & x^9 & x^8 & x^7 & x^6 & x^5 & x^4 & x^3 & x^2 & x
\end{array}
\right]
$$
\end{footnotesize}

\noindent proposed in \cite{Zie} and $L$ (resp. $M$) and $M$ (resp. $N$) as the identity matrices of the appropriate dimensions.
Both of the considered algorithms produce the following Moore-Penrose inverse:

\begin{footnotesize}
$$A^{\dagger}=
\left[
\begin{array}{llllllllllll}
 \frac{x}{x^2+1} & 0 & 0 & 0 & 0 & 0 & 0 & 0 & 0 & 0 & 0 & 0 \\
 \frac{1}{x^2+1} & 0 & 0 & 0 & 0 & 0 & 0 & 0 & 0 & 0 & 0 & 0 \\
 -x & 1 & 0 & 0 & 0 & 0 & 0 & 0 & 0 & 0 & 0 & 0 \\
 0 & -x & 1 & 0 & 0 & 0 & 0 & 0 & 0 & 0 & 0 & 0 \\
 0 & 0 & -x & 1 & 0 & 0 & 0 & 0 & 0 & 0 & 0 & 0 \\
 0 & 0 & 0 & -x & 1 & 0 & 0 & 0 & 0 & 0 & 0 & 0 \\
 0 & 0 & 0 & 0 & -x & 1 & 0 & 0 & 0 & 0 & 0 & 0 \\
 0 & 0 & 0 & 0 & 0 & -x & 1 & 0 & 0 & 0 & 0 & 0 \\
 0 & 0 & 0 & 0 & 0 & 0 & -x & 1 & 0 & 0 & 0 & 0 \\
 0 & 0 & 0 & 0 & 0 & 0 & 0 & -x & 1 & 0 & 0 & 0 \\
 0 & 0 & 0 & 0 & 0 & 0 & 0 & 0 & -x & 1 & 0 & 0 \\
 0 & 0 & 0 & 0 & 0 & 0 & 0 & 0 & 0 & -x & \frac{1}{x^2+1} & \frac{x}{x^2+1}
\end{array}
\right].
$$
\end{footnotesize}
\end{exm}

\begin{exm} \rm
The CPU time needed for the computation of the generalized $LM$-inverse and the weighted Moore-Penrose inverse
(according to Algorithm \ref{algudwa} and the algorithms \ref{algwang}, \ref{algwang1}, respectively) is compared in the next table.
The testing is done on the local machine with the following performances:
Windows edition: Windows Home Edition; Processor: Intel(R) Celeron(R) M \ CPU @ 1.6GHz;
Memory (RAM): 512 MB; System type: 32-bit Operating System; Software: {\ssr MATHEMATICA} 5.2.
Also, in the table there are arranged results obtained on the set of randomly generated test matrices $A_{m \times n}$ and randomly generated symmetric positive definite
matrices $L_{m \times m}$ (resp. $M_{m \times m}$) and $M_{n \times n}$ (resp. $N_{n \times n}$):

\begin{center}
\begin{scriptsize}
\begin{tabular}{|c|c|c|c|}\hline
$m\times n$ & degree &
$\begin{array}{cc}
Algorithm \ \ref{algwang}
\\ A_{MN}^\dag\end{array}$ &
$\begin{array}{cc}
Algorithm \ \ref{algudwa} \\ A_{LM}^\dag\end{array}$  \\
  \hline
  5x6  & 1 &2.265 Seconds          &2.235 Seconds            \\
5x6  & 2 &4.078            &4.063             \\
6x4  & 5 &9.969            &9.625             \\
6x4  & 10 &23.328           &21.11              \\
10x11 &1  &104.109            &103.484        \\
10x11  & 2&192.734             &186.297            \\
11x10  & 1&133.469        &130.125     \\
11x10  & 2&261.359          &227.047         \\
\hline
\end{tabular}
\end{scriptsize}

\medskip
Table 1. The comparison in the efficiency on the set of randomly generated test matrices
\end{center}
\end{exm}

According to the above Table 1 results it is evident that Algorithm \ref{algudwa} produces negligibly better performances with respect to Algorithm \ref{algwang} for all test cases.
This fact is in accordance with the verified equivalence between the algorithms.

\section{Conclusions}

Our primary idea is to show that the computational method for the
generalized $LM$-inverse from \cite{Udwadia,Udwadia2} and the algorithms for the computation of the
weighted Moore-Penrose inverse from \cite{Wang} are equivalent.
The effective algorithm for the computation of the generalized $LM$-inverse is introduced here.
Equivalence of the considered algorithms is proved in the third section by verifying the equivalence of outputs generated by
corresponding algorithmic steps.
This paper not only compares the corresponding algorithms but also compares the performance of two approaches of finding the Moore-Penrose inverse.
In order to compare the efficiency of corresponding algorithms we developed their implementations in the programming
language {\ssr MATHEMATICA}.

\section{Appendix}

Several auxiliary procedures implemented in {\ssr MATHEMATICA} are described at the beginning.

\lstset{commentstyle=\textit}

\begin{footnotesize}
\begin{lstlisting}
TakeCol[A_, k_] := Transpose[Take[Transpose[A], {k}]];
TakeCols[A_, k_] := Transpose[Take[Transpose[A], k]];
TakeElement[A_, i_, j_] := TakeCol[Take[A, {i}], j];
TakeElements[A_, i_, j_] := TakeCols[Take[A, i], j];
T[A_] := Transpose[A];
RandomPoly[n_, prob1_, prob2_, var_] := Module[{S, i, l},
  If [Random[Real, {0, 1}] > prob1, Return[0]];
  S = 0;
  Do[ If [Random[Real, {0, 1}] < prob2, l = 1, l = 0];
    S = S + var^i*Random[Integer, {-10, 10}]*l, {i, 0, n}];
  If [(S == 0) && (prob1 >= 1), S = S + 1];
  Return[S]];
RandomMatrix[m_, n_, deg_, prob1_, prob2_, var_] := Module[{A, i, j},
  A = Table[0, {i,1,m}, {j,1,n}];
  Do[Do[A[[i, j]] = RandomPoly[deg, prob1, prob2, var], {i,1,m}], {j,1,n}];
  Return[A]];
\end{lstlisting}
\end{footnotesize}

Implementation of Algorithm \ref{algudwa} is given by the following function.

\begin{footnotesize}
\begin{lstlisting}
ALMW[A_, M_, N_] :=
  Module[{I, m, n, d, c, tmp, tmp1, u, q, p, a, A1, l, X, N1, N2, n11},
   a = TakeCol[A, 1];  {m, n} = Dimensions[A];
   If [Together[a] === 0*a,
    X = T[a],
    X = Inverse[T[a].M.a].T[a].M ];
   A1 = {};
 Do[I = IdentityMatrix[i - 1];
    a = TakeCol[A, i]; A1 = TakeCols[A, i - 1];
    n11 = TakeElement[N, i, i];  N1 = TakeElements[N, i - 1, i - 1];
    N2 = TakeElements[N, i, i];   l = T[TakeCols[Take[N, {i}], i - 1]];
    d = (IdentityMatrix[m] - A1.X).a // Together;
    p = (I - X.A1).Inverse[N1].l // Together;
    If [Together[d] =!= 0*d,
     b = Inverse[T[d].M.d].T[d].M; b = T[b] // Together,
     tmp = (X.a + p) // Together; q = Join[tmp, {{-1}}];
     u = Append[X, Table[0, {j, m}]];
     b = (T[q].N2.u) Inverse[T[q].N2.q][[1]];
     b = T[b] // Together;
     ];
    tmp1 = X - (X.a - p).T[b] // Together;
    X = Together[Join[tmp1, T[b]]];
    , {i, 2, n}];
   Return[X // MatrixForm]];
\end{lstlisting}
\end{footnotesize}

Implementation of Algorithm \ref{algwang} is obtained by slightly adopting the {\ssr MATHEMATICA} code described in
\cite{Tasic1}.

\begin{footnotesize}
\begin{lstlisting}
AWang[A_, M_, N_] :=
 Module[{I, m, n, d, c, a, A1, l, del, X, tmp, tmp1, N1, n11},
   a = TakeCol[A, 1];
   {m, n} = Dimensions[A];
   If [Together[a] === 0*a,
     X = T[a],
     X = Inverse[T[a].M.a].T[a].M ]; A1 = {};
    Do[ I = IdentityMatrix[i - 1];
     a = TakeCol[A, i];  A1 = TakeCols[A, i - 1];
     d = X.a;   c = Together[a - A1.d];
     n11 = TakeElement[N, i, i];
     N1 = TakeElements[N, i - 1, i - 1];
     l = T[TakeCols[Take[N, {i}], i - 1]];
     tmp = T[d].N1 - T[l];
     tmp1 = (I - X.A1).Inverse[N1].l // Together;
     If [Together[c] =!= 0*c,
       b = Inverse[T[c].M.c].T[c].M;
       b = T[b] // Together,
       del = n11 + T[d].N1.d - (T[d].l + T[l].d) - T[l].tmp1 // Together;
       b = T[Inverse[del].tmp.X] // Together;
      ];
      X = Together[Join[X - (d + tmp1).T[b], T[b]]];
      , {i, 2, n}];
     Return[X]];
\end{lstlisting}
\end{footnotesize}


\begin{thebibliography}{22}

\bibitem{BIG} A. Ben-Israel and  T.N.E. Greville, {\em Generalized inverses: theory and applications},
Second Ed., Springer, 2003.

\bibitem{Campbell} S.L. Campbell and C.D. Meyer, Jr.,
{\em Generalized inverses of linear transformations}, London, Pitman, 1979.

\bibitem{Fan} Y. Fan a and R. Kalaba, {\em Dynamic programming and pseudo-inverses},
Appl. Math. Comput. {\bf 139} (2003), 323–-342.

\bibitem{Gre} T.N.E. Greville,
{\em Some applications of the pseudo-inverse of matrix} SIAM Rev.
{\bf 3} (1960),  15--22.

\bibitem{Itiki} C. Itiki,
{\em Dynamic programming and diagnostic classification},
J. Optim. Theory Appl. {\bf 127} (2005), 579–-586.


\bibitem{Kurmayya} T. Kurmayya and K.C. Sivakumar,
{\em Moore-Penrose inverse of a Gram matrix and its nonnegativity},
J. Optim. Theory Appl. {\bf 139} (2008), 201–-207.

\bibitem{Layton} J.B. Layton,
{\em Efficient direct computation of the pseudo-inverse and its gradient},
Internat. J. Numer. Methods Engrg. {\bf 40} (1997), 4211--4223.

\bibitem{Mohi} S. Mohideen and V. Cherkassky,
{\em On recursive calculation of the generalized inverse of a matrix},
ACM Trans. Math. Software {\bf 17} (1991), 130–-147.

\bibitem{Petkovic} M.D. Petkovi\' c, P.S. Stanimirovi\' c and M.B. Tasi\' c,
{\em Effective partitioning method for computing weighted Moore–Penrose inverse},
Comput. Math. Appl. {\bf 55} (2008), 1720–-1734.

\bibitem{Stanimirovic3} M.D. Petkovi\' c and P.S. Stanimirovi\' c,
{\em Symbolic computation of the Moore-Penrose inverse using partitioning method},
Int. J. Comput. Math. {\bf 82} (2005), 355--367.

\bibitem{Phoho} P. Phohomsiri, B. Han,
{\em An alternative proof for the recursive formulae for computing the Moore–Penrose M-inverse of a matrix},
Appl. Math. Comput. {\bf 174} (2006), 81-–97.

\bibitem{Sivakumar} K. C. Sivakumar,
{\em Proof by verification of the Greville/Udwadia/Kalaba formula for the Moore-Penrose inverse of a matrix},
J. Optim. Theory Appl. {\bf 131} (2006), 307–-311.

\bibitem{Stanimirovic2} P.S. Stanimirovi\' c and M.B. Tasi\' c,
{\em Partitioning method for rational and polynomial matrices},
Appl. Math. Comput. {\bf 155} (2004), 137--163.

\bibitem{Tasic1} M.B. Tasi\' c, P.S. Stanimirovi\'c, M.D. Petkovi\'c,
{\em Symbolic computation of weighted Moore-Penrose inverse using partitioning method},
Appl. Math. Comput. {\bf 189} (2007), 615--640.

\bibitem{Tasic2} M.B. Tasi\' c, P.S. Stanimirovi\'c,
{\em Symbolic and recursive computation of different types of generalized inverses},
Appl. Math. Comput. {\bf 199} (2008), 349--367.

\bibitem{Udwadia} F.E. Udwadia and P.Phohomsiri,
{\em Generalized $LM$-inverse of a matrix augmented by a column vector}, Appl. Math. Comput.
{\bf 190} (2007), 999--1006.

\bibitem{Udwadia2} F.E. Udwadia and P.Phohomsiri, {\em Recursive Formulas for Generalized LM-Inverse of a Matrix},
J. Optim. Theory Appl. {\bf 131} (2007), 1--16.

\bibitem{Udwadia3} F. E. Udwadia and P. Phohomsiri,
{\em Recursive Determination of the Generalized Moore–Penrose M-Inverse of a Matrix},
J. Optim. Theory Appl. {\bf 127} (2005), 639-–663.

\bibitem{Udwadia2N} F.E. Udwadia and R.E. Kalaba, {\em An Alternative Proof for Greville's Formula},
J. Optim. Theory Appl. {\bf 94} (1997), 23-28.

\bibitem{Udwadia3N} F.E. Udwadia and R.E. Kalaba, {\em Analytical Dynamics: A New Approach},
Cambridge University Press, Cambridge, England, 1996.

\bibitem{UdwadiaN} F.E. Udwadia and R.E. Kalaba, {\em A Unified Approach for the Recursive Determination
of Generalized Inverses}, Comput. Math. Appl. {\bf 37} (1999), 125-130.

\bibitem{UdwadiaJOTA} F.E. Udwadia and R.E. Kalaba, {\em General forms for the Recursive Determination
of Generalized Inverses: Unified approach}, J. Optim. Theory Appl. {\bf 101} (1999), 509--521.

\bibitem{Wang} G.R. Wang and Y.L.Chen,
{\em A recursive algorithm for computing the weighted Moore-Penrose inverse $A_{MN}^{\dagger}$},
J. Comput. Math. {\bf 4} (1986), 74--85.

\bibitem{Wang1} G.R. Wang,
{\em A new proof of Greville's method for computing the weighted M-P inverse},
J. Shangai Teach. Univ., Nat. Sci. Ed. {\bf 3} (1985), 32--38.

\bibitem{Zie} G. Zielke,
{\em Report on test matrices for generalized inverses}, Computing
{\bf 36} (1986), 105--162.

\end{thebibliography}
\end{document}